\documentclass[12pt,a4paper]{article}

\usepackage[dvips]{graphicx}
\usepackage{amsmath}
\newcommand{\E}{\; {\rm Exp} }

\def\a{\alpha_s}
\def\A{{\hat\alpha}_s}

\begin{document}
\begin{center}
         {\large \bf Improved valence PDFs 
at leading order via connection 
formulas between the different QCD orders 
}
 \end{center}
 \begin{center}
 \vskip 1cm
 {\large O.\,Yu.\,Shevchenko}\footnote{E-mail address: shev@mail.cern.ch},
 {\large R.\,R.\,Akhunzyanov}\footnote{E-mail address: axruslan@mail.ru}
 \vspace{0.5cm}

 {\it Joint Institute for Nuclear Research}
 \end{center}

\begin{abstract}

The formulas directly connecting
parton distribution functions (PDFs) 
at the
leading (LO) and the next to leading (NLO) QCD orders are applied to both
unpolarized and polarized valence PDFs and produce
improved LO results on these quantities
without exclusion from the analysis the data coming from the small and moderate $Q^2$ region.
It is shown that derived in this way and presented in the paper improved LO parametrizations of the valence PDFs
could be obtained within the standard LO analysis
by using only  the data produced  
at very high $Q^2$ values (that is hardly possible in reality) and they
strongly differ from the original LO
parametrizations of these quantities, 
obtained by using the most of data 
produced at small and moderate $Q^2$ values.
It is argued that LO results on PDFs can be significantly improved even if the respective NLO results
are still unavailable.

\end{abstract}
\begin{flushleft}
\end{flushleft} 

QCD analysis of the experimental data at LO is a very useful tool for the
understanding of hadronic high energy reactions. This is due
to its simplicity and comprehensiveness as well as to its
universality, i.e. applicability to any hadronic process. 
Of importance is also (and it seems to be the main advantage of LO)
that in contrast to PDFs and fragmentation functions (FFs) at higher QCD orders   
these quantities at LO  not only can be used as an input for calculation (in the chosen factorization scheme)
of asymmetries and cross sections of 
different processes, but being scheme independent have 
the physical meaning themselves,
allowing the probabilistic interpretation  and, thereby, the possibility of qualitative physical predictions.

All these reasons make LO analysis very needed even today and many modern collaborations produce
namely LO results on PDFs/FFs (see, for instance, \cite{compass2010, HERMES}).
Besides, it is of importance also, that until now only LO analysis is still possible (see \cite{Sivers1, Sivers2}
and references therein) 
for asymmetries giving access to 
such intensely studied nowadays 
PDFs as transverse momentum distributions (TMDs). 
At the same time, as we will see below, the inclusion of the data produced at small and moderate $Q^2$ in the standard LO
analysis is really dangerous for the LO results:
as we will see below,  
the difference between the LO results on PDFs with and without inclusion of the small and moderate $Q^2$ data
into analysis is significant and is not covered by the error band. 
However, now did appear a good possibility to produce the correct LO results on PDFs/FFs without lost
of events 
collected in experiment at small and moderate $Q^2$, which 
make up the overwhelming majority of events available to any experiment.

Recently \cite{CF} the formulas directly connecting
parton distribution functions and fragmentation functions at 
the next to leading order of QCD with the same quantities at the leading 
order were derived (see Eqs. (15), (16) in Ref. \cite{CF}). 
To obtain these formulas 
only the DGLAP evolution equations and the asymptotic 
condition 
that PDFs (FFs) at different QCD orders become the same in the Bjorken limit were used
as an input.
Due to universality of this input the obtained connection formulas are also universal,  
i.e. they are valid for any kind of PDFs (FFs) we deal with, differing only in the 
respective splitting functions entering there. Moreover, 
operating in the same way as in Ref.  \cite{CF}
one can also establish the connection of 
PDFs (FFs) at LO (as well as at NLO) with these quantities at any higher 
QCD order (NNLO, NNNLO, $\ldots$) (will be published elsewhere).

In this paper we focus on the (both unpolarized and polarized) valence PDFs. 
The non-singlet connection formulas, Eq. (16) in Ref. \cite{CF}, for the unpolarized ${ q}_V\equiv  q- {\bar q}$
and polarized  $\Delta q_V\equiv  \Delta q- \Delta \bar q$  
(helicity PDFs)
valence PDFs  
look as
\begin{eqnarray}
\label{LOtoNLO}
&q_V(Q^2,x)=    
\left[\delta(1-x)+ \frac{\a(Q^2)}{2\pi}\left(\frac{\beta_1}{\beta_0^2} 
P^{(0)}(x) -\frac{2}{\beta_0} P^{(1)}(x) 
\right) \right] \nonumber \\ 
&\otimes
\E \left(- \frac{2}{\beta_0}\ln\frac{\a(Q^2)}{\A(Q^2)}\, P^{(0)}(x) \right) 
\otimes
{\hat q}_V(Q^2,x),
\end{eqnarray}
and the same 
for the valence helicity PDFs $\Delta q_V\equiv  \Delta q- \Delta \bar q$ with the 
replacements\footnote{
All over the paper we will use the most often used $\overline {\rm MS}$ scheme for the quantities at NLO. 
The respective unpolarized $
P_V^{(1)}(x)$ 
and polarized   
$
\Delta P_V^{(1)}(x)$
splitting functions can be found, for instance, in Refs. 
\cite{FurmanskiPetronzio1980} and \cite{Vogelsang}.
} 
$q_V ({\hat q}_V) \to \Delta q_V (\Delta {\hat q}_V)$, $P_V \to \Delta P_V$.
Here the notation of Ref. \cite{CF} 
\begin{equation}
A \bigl|_{LO}\equiv \hat A, \quad  A \bigl|_{NLO}\equiv A,
\end{equation}
for any quantity $A$ 
at
LO and NLO is used, 
the convolution ($\otimes$) is defined in standard way
\begin{equation}
(A\otimes B)(x)= \int^1_0 dx_1\int^1_0dx_2\, \delta(x-x_1x_2)A(x_1) B(x_2) = \int^1_x \frac{dy}{y} A(\frac{x}{y}) B(y),
\nonumber
\end{equation}
while the generalized exponent $\E \bigl( \epsilon\, P_V^{(0)}(x) \bigl)$ in Eq. (\ref{LOtoNLO})
is defined as 
a series 
\begin{equation}
\label{series}
\E \left (\epsilon\, P_V^{(0)}(x)\right )
= \delta(1-x) + \epsilon\,  P_V^{(0)}(x)+(\epsilon^2/ 2!)\,P_V^{(0)}(x) 
\otimes P_V^{(0)}(x) + \ldots,
\end{equation}
and analogously for $\E \bigl( \epsilon\, \Delta P_V^{(0)}(x) \bigl)$.
The parameter  $\epsilon \equiv
(-2/\beta_0)\ln(\a/\A)$
is very small even at the minimal really available $Q^2$ values (the lower boundary of the experimental cut on 
$Q^2$ is usually about $1\text{ GeV}^2$), so 
one can achieve very good accuracy keeping only few first terms in 
this expansion. Within this paper we will keep only two terms in the expansions like Eq. (\ref{series}),
because $O(\epsilon^2)$ corrections are really negligible 
at the reference scale $Q_{ref}^2=10\text{ GeV}^2$ which we use in the paper.

Let us make an important remark concerning the validity of Eq. (\ref{LOtoNLO}), which is just the particular
realisation of non-singlet connection formula (16) in Ref. \cite{CF}. One can be rather confused by the application of 
evolution up to infinite point and inverse evolution to initial point applied at the derivation of the 
connection formulas (15), (16) in Ref. \cite{CF}.
However, of importance is that to understand the correctness of Eqs. (15), (16) in Ref. \cite{CF}, 
the history 
of obtaining these formulas
is not especially important,
because one can just forget for a moment all delicate moments of their derivation and 
{\it directly check} their validity (see footnote 3 in Ref.  \cite{CF}). In particular,
using   
the obvious relation $Q^2d \bigl[\E \left((2/\beta_0) \ln\A \, {P_V^{(0)}} \right)
\otimes {\hat q}_V \bigl]/dQ^2=0$ 
one can immediately check
that r.h.s. of Eq. (\ref{LOtoNLO}) indeed satisfies
the NLO DGLAP evolution equation 
\begin{equation}
\label{dglap}
Q^2 d q_V(Q^2,x)/dQ^2=
(\a/2\pi)[P_V^{(0)}(x) + (\a/2\pi) P_V^{(1)}(x)+ O(\a^2)]\otimes 
q_V(Q^2,x),
\end{equation}
if only ${\hat q}_V$ satisfies LO DGLAP equation  
$Q^2 d {\hat q}_V(Q^2,x)/dQ^2=
(\a/2\pi) P_V^{(0)}(x) \otimes  {\hat q}_V(Q^2,x)$.
Besides, one can see that $q$ and $\hat q$ connected by Eq. (\ref{LOtoNLO}) indeed coincide
in the Bjorken limit, i.e., satisfy the asymptotic condition Eq. (11) in Ref. \cite{CF}, which is the key point for the derivation 
of connection formulas.

Let us now proceed with an {\it important statement}:
applying the connection formulas like Eq. (\ref{LOtoNLO}) one should avoid the naive substitution of the 
existing LO parametrizations to r.h.s.
of these formulas. 
The point is that all LO parametrizations of PDFs (FFs) existing today  are obtained using the
LO theoretical expressions (i.e., excluding all QCD corrections to improved QPM) for the 
cross sections and asymmetries measured in experiment 
(actually including {\it all QCD corrections})
at all, even very small (about $1\text{ GeV}^2$) available 
$Q^2$ values. 
At the same time, as we will see below, 
this is dangerous
procedure and leads to the strong distortion of the PDFs behaviour at LO. 
We will see, that such procedure of LO analysis can produce correct (realistic) LO PDFs (FFs) only in very high $Q^2$ 
region 
and only PDFs (FFs) obtained there should then be 
evolved via DGLAP to the small and moderate $Q^2$ region.
Regretfully,  
the most of statistics is available just at small and moderate $Q^2$, so that this way 
of action 
seems to be hardly feasible. 
However, as it was mentioned above,
now there is a good possibility to approach the correct 
PDFs/FFs at LO without any restrictions on $Q^2$ for the applied data. 
To this end one should just inverse 
the connection formulas 
(Eqs. (15), (16) in Ref. \cite{CF}),
thereby expressing LO quantities through extracted from the data NLO ones.
Certainly, PDFs (FFs) at LO obtained in this way 
are much more realistic since one believes that
NLO theoretical 
expressions for asymmetries and cross sections well approximate their measured values even at low $Q^2$. 
Thus, from now on we will call such PDFs/FFs the ``improved PDFs/FFs'' at LO.         
Later we will argue (see a numerical experiment below) that such improved PDFs/FFs can be reproduced 
within the standard procedure of LO analysis only by excluding the dangerous data produced at small
and moderate $Q^2$.

Let us first 
consider very important case of unpolarized parton densities  
and obtain the improved non-singlet unpolarized valence PDFs ${\hat q}_V\equiv \hat q-\hat {\bar q}$ at LO 
via substitution of some modern NLO parametrization of ${ q}_V$ into
the inverse of  Eq. (\ref{LOtoNLO}) connection formula
\begin{eqnarray}
\label{NLOtoLO}
&{\hat q}_V(Q^2,x)=    
\left[\delta(1-x)- \frac{\a(Q^2)}{2\pi}\left(\frac{\beta_1}{\beta_0^2} 
P^{(0)}(x) -\frac{2}{\beta_0} P^{(1)}(x) 
\right) \right] \nonumber \\ 
&\otimes
\E \left( \frac{2}{\beta_0}\ln\frac{\a(Q^2)}{\A(Q^2)}\, P^{(0)}(x) \right) 
\otimes
q_V(Q^2,x),
\end{eqnarray}
and then compare the obtained result with the respective original LO parametrization of ${\hat q}_V$.
We choose here the modern and widely used \cite{mstw} MSTW2008 parametrization 
both for NLO and LO unpolarized PDFs.
 
\begin{figure}
\center
\includegraphics[height=5.5cm]{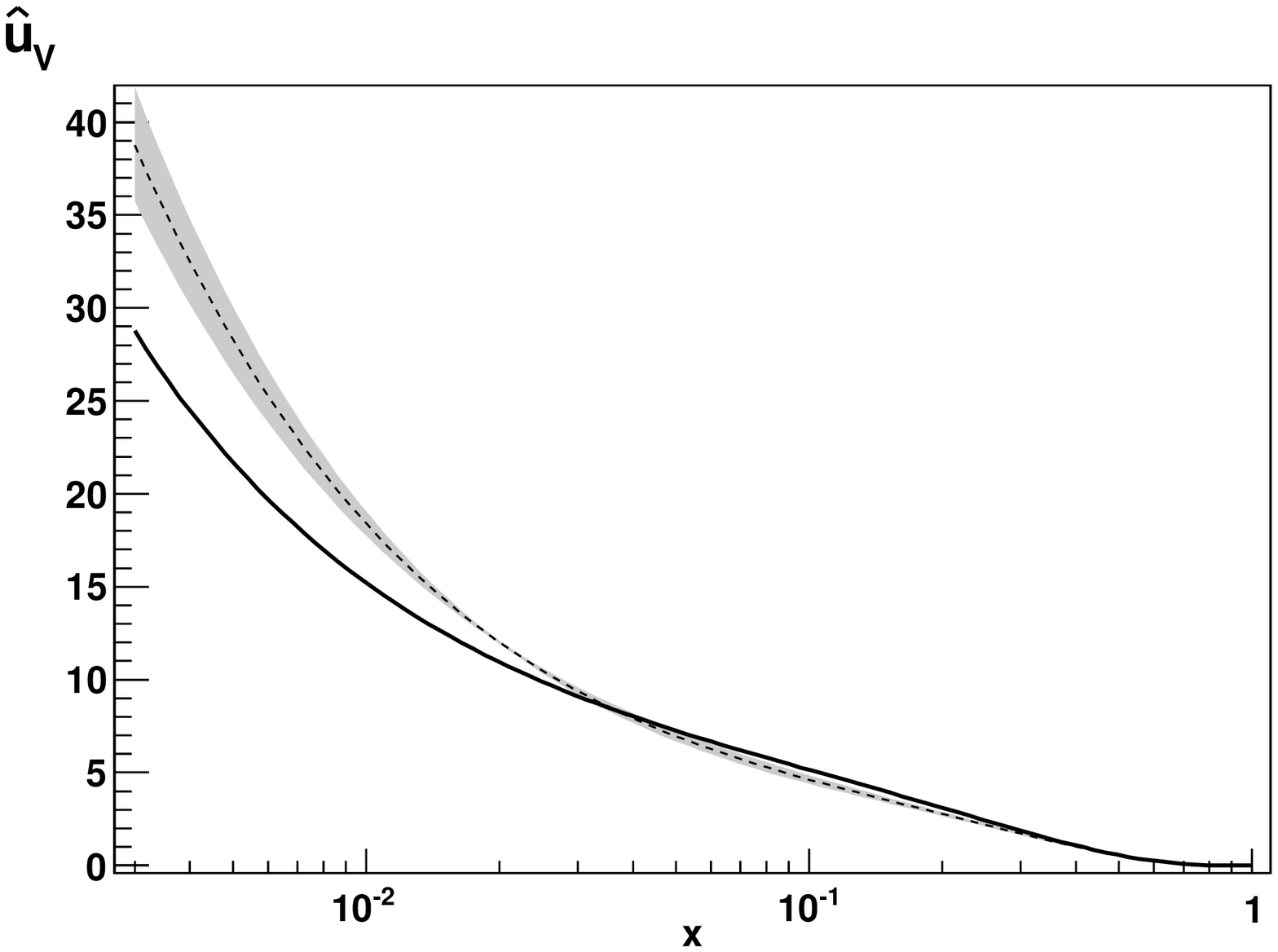}
\includegraphics[height=5.5cm]{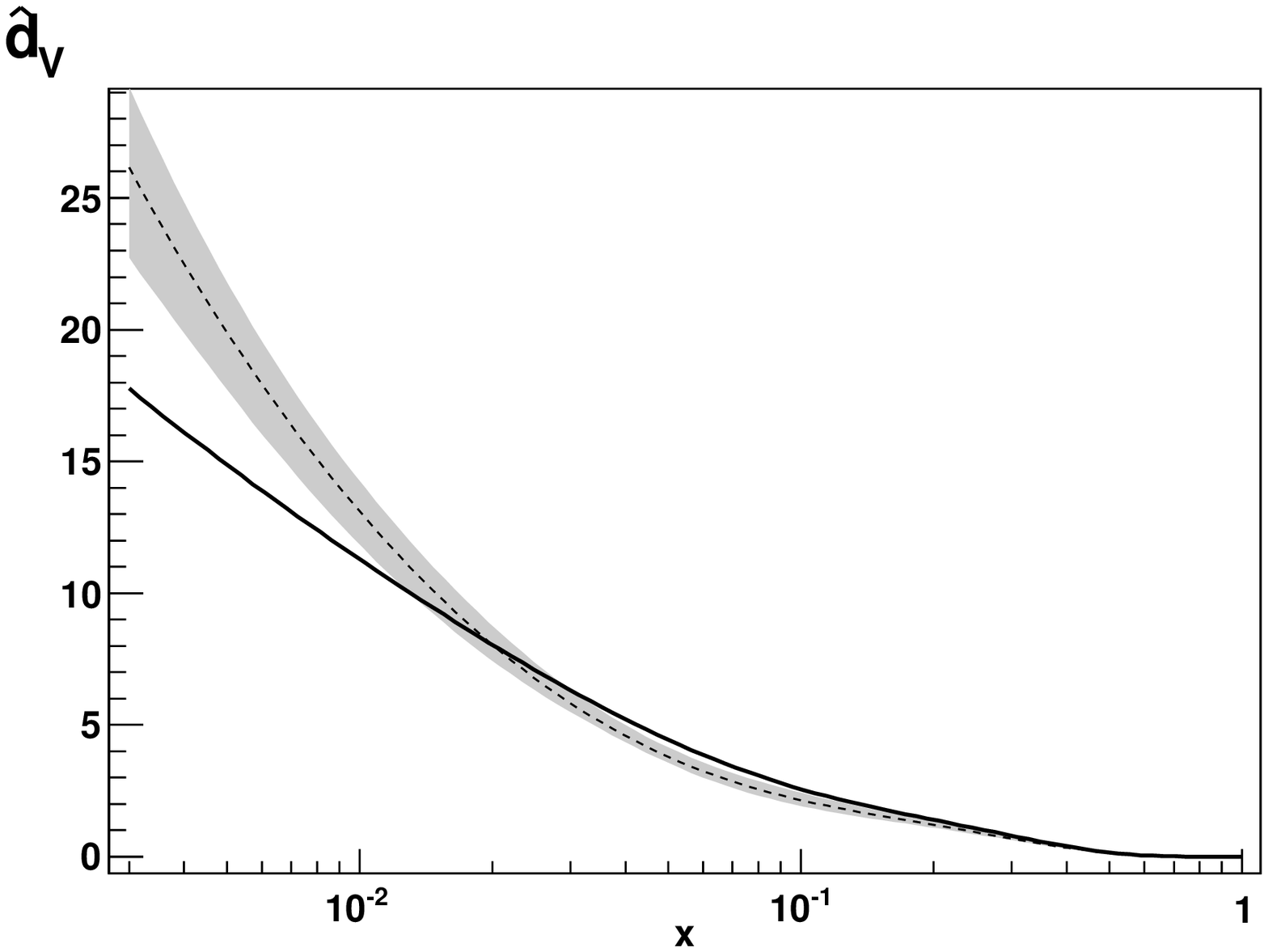}
\caption{\footnotesize Results on ${\hat u}_V$ (left) and ${\hat d}_V$ (right) versus $x$ at  
$Q_{ref}^2=10\text{ GeV}^2$.
The dashed line is MSTW2008 LO parametrization \cite{mstw}. 
the gray area is the corresponding error band.
The bold solid line corresponds to improved MSTW2008LO parametrization of ${\hat u}_V$ and ${\hat d}_V$, which is  
obtained by using the  MSTW2008 NLO parametrization in r.h.s. of inverse connection 
formula Eq. (\ref{NLOtoLO}) 
}
\label{fig:x_lo_uv_Q2_10}
\end{figure}

Improved MSTW2008 LO parametrization of  ${\hat q}_V$  obtained via the inverse connection 
formula (\ref{NLOtoLO})   
is presented in Fig. \ref{fig:x_lo_uv_Q2_10} in comparison with the original
MSTW2008 LO  parametrization of  ${\hat q}_V$ . 
One can see that the improved LO parametrization 
essentially differs from the respective LO parametrization 
on  ${\hat q}_V$ obtained in standard way, and the difference is not covered by error band.

To better realize what happens when one performs LO analysis of the data let us 
perform the {\it numerical experiment}
and prepare the artificial 
``ideal'' polarized semi-inclusive DIS data on pion production.
Namely, we 
consider as such ``pseudo-experimental'' input 
the integrated over cut in $z$ proton and deuteron difference asymmetries\footnote{
These asymmetries is an ideal tool for our research, because in contrast to all other 
known asymmetries (and cross sections) they contain at LO nothing except the valence
PDFs. The similar object $F_2^{\pi^+}- F_2^{\pi^-}$ in unpolarized semi-inclusive DIS is not so convenient for our purposes,
because it involves pion FFs already at LO.  
}
$A_{p(d)}^{``exp"}{\bigl |}_Z$, 
constructed
by substitution of the known\footnote{We will use MSTW2008 \cite{mstw} parametrization of $q$, DSSV \cite{DSSV} or GRSV \cite{GRSV} parametrizations of $\Delta q$ and AKK \cite{AKK} parametrization of FFs.} 
NLO parametrizations of $u_V$, $d_V$,
$\Delta u_V$, $\Delta d_V$ and  
difference of favored and unfavored pion FFs $D_1-D_2$ into the theoretical
expressions for these asymmetries at NLO (see Eqs. (6-10) in Ref. \cite{shev}). 
Then, we extract $\Delta {\hat u}_V$ and  $\Delta {\hat d}_V$ at LO using these constructed asymmetries
$A_{p(d)}^{``exp"}{\bigl |}_Z$ as a  ``pseudo-experimental'' input in the LO equations (all notation is just as in Ref. \cite{shev})
\begin{equation}
\label{LO_pseudo}
A_{p}^{``exp"}{\bigl |}_Z(x,Q^2)  = (1+R)\frac{
4\Delta {\hat u}_V-\Delta {\hat d}_V}
{4{\hat u}_V-{\hat d}_V}, \,\,
A_{d}^{``exp"}{\bigl |}_Z(x,Q^2)  = (1+R)(1-\frac{3}{2}\omega_D)\frac{ 
\Delta {\hat u}_V+\Delta {\hat d}_V}
{{\hat u}_V+{\hat d}_V},
\end{equation}
i.e., we perform the LO analysis analogous to one of COMPASS \cite{compass2008}, but with the replacement
of the real data on $A_{p(d)}^{exp}{\bigl |}_Z$ by the prepared pseudo-data on $A_{p,(d)}^{``exp"}{\bigl |}_Z$.
The obvious advantage of the latter is that they exist in the widest range in $Q^2$, up to $10^6\text{ GeV}^2$
in accordance with the upper $Q^2$ boundary for parametrizations which we use for the pseudo-data preparation.
Making use of this advantage we can now comprehensively investigate the influence of $Q^2$ on the produced with 
Eqs.~(\ref{LO_pseudo}) LO results.  
Namely, we now have a possibility to obtain and compare the LO result on $\Delta {\hat q}_V$ 
at the same reference scale ($Q_{ref}^2=10\text{ GeV}^2$ here)
with and without involving the dangerous data at small and moderate $Q^2$ into the analysis.

\begin{figure}
\center 
\includegraphics[height=6.0cm]{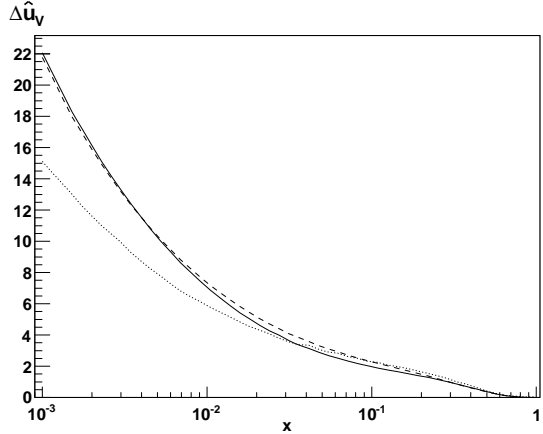}
\caption{ \footnotesize
$\Delta {\hat u}_V$ at  $Q_{ref}^2=10\text{ GeV}^2$ versus $x$.
Dashed line corresponds to the original GRSV LO parametrization.
Solid line corresponds to 
direct solution of the system (\ref{LO_pseudo}) with respect to $\Delta {\hat q}_V$, 
where asymmetries $A_{p,(d)}^{``exp"}{\bigl |}_Z$ are taken at $Q_{ref}^2=10\text{ GeV}^2$  and MSTW2008 LO parametrization
of  ${\hat q}_V$ is used in denominators.
Asymmetries  $A_{p,(d)}^{``exp"}{\bigl |}_Z$ are calculated substituting  
GRSV NLO parametrization of polarized PDFs, 
MSTW2008 NLO parametrization of unpolarized PDFs,
and AKK parametrization of FFs 
into the respective NLO equations (Eqs. (6-10) in Ref. \cite{shev}).
Dotted line corresponds to $\Delta {\hat q}_V$ 
which are obtained in the same way but 
at extremely high $Q^2=10^6\text{ GeV}^2$ and with improved MSTW2008 LO parametrization
of ${\hat q}_V$
(bold solid lines in Fig. \ref{fig:x_lo_uv_Q2_10}) in denominators of Eqs. (\ref{LO_pseudo}).
The obtained in this way $ {\hat q}_V$ are then evolved  
via DGLAP equations from $Q^2=10^6\text{ GeV}^2$ to $Q_{ref}^2=10\text{ GeV}^2$.
}
\label{fig:grsv_lo_qv}
\end{figure}

\begin{figure}
\center
\includegraphics[height=6.0cm]{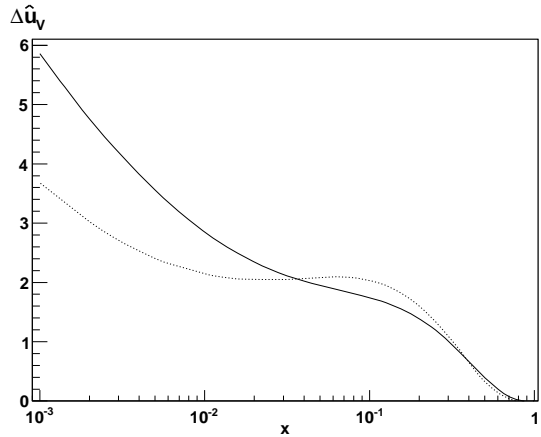}
\caption{\footnotesize
$\Delta {\hat u}_V$ at  $Q_{ref}^2=10\text{ GeV}^2$ versus $x$.
Solid line corresponds to 
direct solution of the system (\ref{LO_pseudo}) with respect to $\Delta {\hat q}_V$, 
where asymmetries $A_{p,(d)}^{``exp"}{\bigl |}_Z$ are taken at $Q_{ref}^2=10\text{ GeV}^2$  and MSTW2008 LO parametrization
on  ${\hat q}_V$ is used in denominators.
Asymmetries  $A_{p,(d)}^{``exp"}{\bigl |}_Z$ are calculated substituting  
DSSV NLO parametrization of polarized PDFs, 
MSTW2008 NLO parametrization of unpolarized PDFs,
and AKK parametrization of FFs 
into the respective NLO equations (Eqs. (6-10) in Ref. \cite{shev}).
Dotted line corresponds to $\Delta {\hat q}_V$ 
which are obtained in the same way but 
at extremely high $Q^2=10^6\text{ GeV}^2$ and with improved MSTW2008 LO parametrization
of ${\hat q}_V$
(bold solid lines in Fig. \ref{fig:x_lo_uv_Q2_10}) in denominators of Eqs. (\ref{LO_pseudo}).
The obtained in this way $ {\hat q}_V$ are then evolved  
via DGLAP equations from $Q^2=10^6\text{ GeV}^2$ to $Q_{ref}^2=10\text{ GeV}^2$.
}
\label{fig:dssv_lo_qv}
\end{figure}

The respective results on $\Delta {\hat u}_V$ are presented in Figs. \ref{fig:grsv_lo_qv} and \ref{fig:dssv_lo_qv} (the 
results on  $\Delta {\hat d}_V$ are analogous),
where, besides the modern parametrization DSSV of $\Delta q$ (Fig. \ref{fig:dssv_lo_qv}), we use 
also older GRSV parametrization as an input in Eqs. (\ref{LO_pseudo}) (Fig. \ref{fig:grsv_lo_qv}), 
because the latter has the LO version, which is useful for comparison.

Dashed line in  Fig. \ref{fig:grsv_lo_qv} corresponds to the original GRSV LO parametrization, while solid and dotted lines 
in  Figs. \ref{fig:grsv_lo_qv}, \ref{fig:dssv_lo_qv}
correspond to 
direct extraction of  $\Delta {\hat q}_V$ with Eqs. (\ref{LO_pseudo}) (direct 
solution of the system (\ref{LO_pseudo}) with respect to $\Delta {\hat u}_V$ and $\Delta {\hat d}_V$),
where
asymmetries  $A_{p,(d)}^{``exp"}{\bigl |}_Z$ are calculated substituting  
GRSV NLO (Fig. \ref{fig:grsv_lo_qv}) or DSSV NLO (Fig. \ref{fig:dssv_lo_qv}) parametrizations of polarized PDFs,  
MSTW2008 NLO parametrization of unpolarized PDFs, 
and AKK parametrization of FFs 
into the respective NLO equations (Eqs. (6-10) in Ref. \cite{shev}).
At the same time, very important difference between  $\Delta {\hat q}_V$ presented by solid and dotted curves 
is that the first is obtained
with  asymmetries $A_{p,(d)}^{``exp"}{\bigl |}_Z$ in Eqs. (\ref{LO_pseudo}) taken 
directly
at $Q_{ref}^2=10\text{ GeV}^2$ , while the second
is obtained with
asymmetries  $A_{p,(d)}^{``exp"}{\bigl |}_Z$ taken at extremely high $Q^2=10^6\text{ GeV}^2$ 
and only then are evolved to the reference scale $Q_{ref}^2=10\text{ GeV}^2$.
Besides, of importance is that instead of 
the original MSTW2008 LO parametrization of ${\hat q}_V$ used for construction of solid lines in Figs. \ref{fig:grsv_lo_qv} and  \ref{fig:dssv_lo_qv},  
the respective improved LO parametrization (given by bold solid lines in Fig. \ref{fig:x_lo_uv_Q2_10}) 
is used in denominators of Eqs. (\ref{LO_pseudo}) for construction of the dotted lines. 
Indeed, it is clear that to produce the completely corrected (improved) LO results on $\Delta{\hat q}_V$ 
one should completely exclude the influence of small $Q^2$ on the analysis with Eq. (\ref{LO_pseudo}).
To this end it is not sufficient to restrict the pseudo-data in l.h.s. of Eqs. (\ref{LO_pseudo}) by the high $Q^2$ values
and one should\footnote{
Otherwise, one implicitly involves in the LO results on $\Delta {\hat q}_V$ 
the small and moderate $Q^2$ data used to produce
the original LO parametrization on ${\hat q}_V$ and obtains some intermediate 
results: 
our calculations show that the respective line lies between dotted and solid (dashed) lines in Fig. \ref{fig:grsv_lo_qv}.}  
also properly correct/improve the denominators in Eqs. (\ref{LO_pseudo}).

Comparing solid and dashed curves in Fig. \ref{fig:grsv_lo_qv} one can see that they are very close to each other, 
which is not surprising,
since the vast majority of data used for production of the original GRSV LO parametrization (as well as for 
production of any other known parametrization) 
correspond to small and moderate  $Q^2$ values close to $10\text{ GeV}^2$. In contrast,  
both of these curves in Fig. \ref{fig:grsv_lo_qv}, as well as solid line in Fig. \ref{fig:dssv_lo_qv} 
strongly differ from the respective dotted lines, corresponding to the ``pure'' LO results on 
$\Delta {\hat q}_V(10\text{ GeV}^2,x)  $,
which are free from the involvement of small $Q^2$ data in the analysis.  

We will see now that these ``pure'' LO results (which within the standard approach could be obtained only completely
excluding small and moderate $Q^2$ data from the analysis) can be reproduced by using the connection formula 
between LO and NLO QCD orders. 

\begin{figure}
\center
\includegraphics[height=6.0cm]{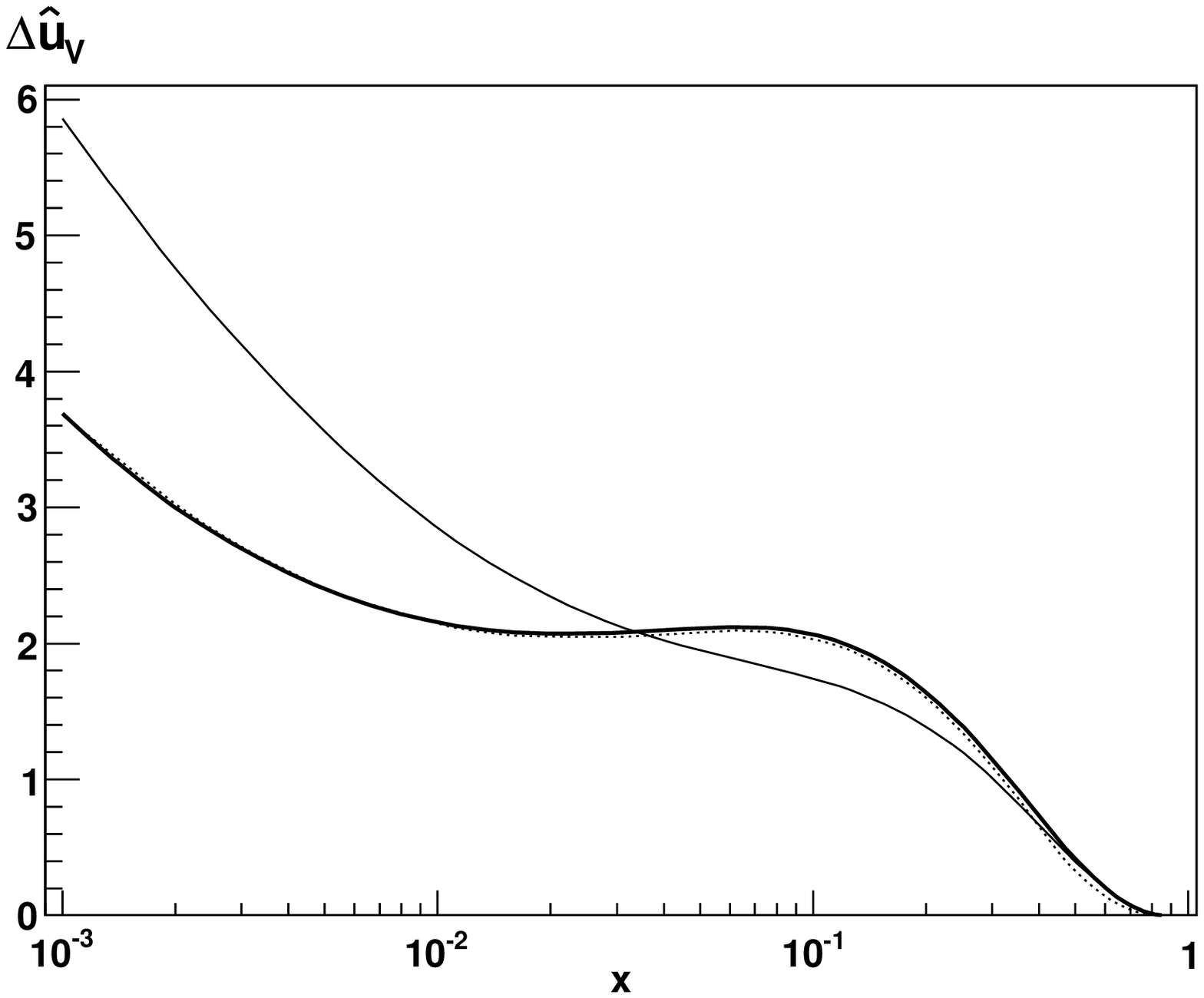}
\includegraphics[height=6.0cm]{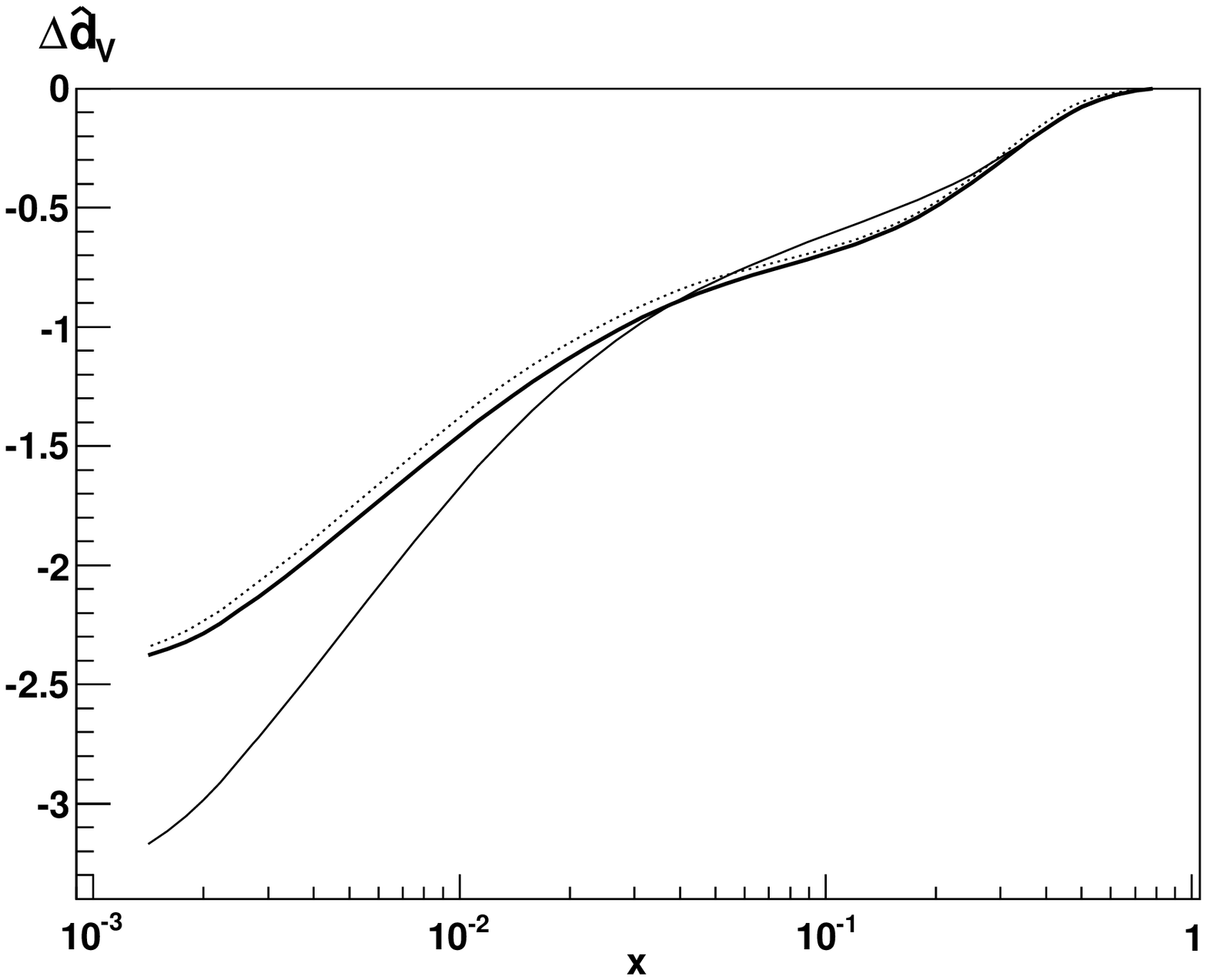}
\caption{
\footnotesize
LO valence helicity PDFs $\Delta {\hat u}_v$ (left) and $\Delta {\hat d}_v$ (right) at  $Q_{ref}^2=10\text{ GeV}^2$  
obtained in different ways.
Bold solid line corresponds to improved LO parametrization, which is obtained
from DSSV NLO parametrization via inverse connection formula (\ref{NLOtoLO_pol}).
Thin solid and dotted lines are just the same as in Fig. \ref{fig:dssv_lo_qv}.  
}
\label{fig:dssv_lo_qv_plus_our}
\end{figure}

Thus, let us 
apply the inverse of Eq. (\ref{LOtoNLO}) connection formula
\begin{eqnarray}
\label{NLOtoLO_pol}
&\Delta {\hat q}_V(Q^2,x)=    
\left[\delta(1-x)- \frac{\a(Q^2)}{2\pi}\left(\frac{\beta_1}{\beta_0^2} 
\Delta P^{(0)}(x) -\frac{2}{\beta_0} \Delta P^{(1)}(x) 
\right) \right] \nonumber \\ 
&\otimes
\E \left( \frac{2}{\beta_0}\ln\frac{\a(Q^2)}{\A(Q^2)}\, \Delta P^{(0)}(x) \right) 
\otimes
\Delta q_V(Q^2,x),
\end{eqnarray}
written in terms of the valence helicity PDFs. 
Substituting in r.h.s. of (\ref{NLOtoLO_pol}) 
the modern DSSV NLO parametrization of $\Delta {q}_V$ (used for production of Fig. \ref{fig:dssv_lo_qv}) 
taken at the reference
scale  $Q^2_{ref}=10\text{ GeV}^2$ 
we obtain 
the respective 
LO results on $\Delta {\hat q}_V$ at the same $Q^2_{ref}$.
The results on $\Delta {\hat q}_V(10\text{ GeV}^2,x)$ obtained in this way are presented in 
Fig. \ref{fig:dssv_lo_qv_plus_our} 
by the bold solid lines and are compared with the respective results, obtained 
earlier by the use of Eqs. (\ref{LO_pseudo}) and presented in 
Fig. \ref{fig:dssv_lo_qv}.
Comparing dotted and bold solid lines one can see that while
they strongly differ from the thin solid line 
directly obtained via Eq. (\ref{LO_pseudo}) at small $Q^2_{ref}=10\text{ GeV}^2$,  
they are in excellent agreement with each other.
Moreover, they 
practically merge with each other.
Thus, one can conclude that to reach improved LO valence PDFs 
there is no need to restrict the analysis to use of data produced only at extremely high
$Q^2$ values (that is hardly possible in reality): the inverse connection formulas             
give the direct access to the same improved LO PDFs 
without any additional\footnote{Only the standard cut $Q^2>1\text{ GeV}^2$, providing pQCD applicability, 
should be applied.} restrictions on the $Q^2$ range from which the data comes. 
In contrast, the
standard procedure of LO analysis can produce these improved PDFs (FFs)
at LO only from data in very high $Q^2$ 
region and only after that they should be 
evolved with DGLAP to the lower  $Q^2$ values. However, regretfully,  
the most of statistics is available just at small and moderate $Q^2$, so that the such a way 
of action is realizable in our simple model but seems to be hardly possible in reality.

Let us now briefly discuss 
some aspects of application of the results on improved PDFs at LO. 
From this point of view of especial importance is the considered above unpolarized case (see Fig. \ref{fig:x_lo_uv_Q2_10}), 
because 
unpolarized parton densities enter the denominators of all asymmetries studied 
nowadays, and, in particular, of the asymmetries giving access to 
such important and  attracting today the great attention 
PDFs as transverse momentum distributions (TMDs), 
for which only LO analysis is still possible (see \cite{Sivers1, Sivers2} and references therein).
Notice, that 
even in this case there is a good possibility to partially (and significantly) improve 
LO results on these quantities without knowledge of the respective NLO results as an input 
in the connection formulas.
Such improvement can be achieved by 
performing the standard
LO analysis of the respective measured asymmetries, 
but using in their denominators the improved 
unpolarized PDFs at LO instead of the usual LO parametrizations.

Let us see how large can be impact on the LO results
only due to improvement of the denominators in the 
asymmetries analysed at LO in standard way at small $Q^2$ values.
To this end we return to the considered above numerical experiment 
based on application of Eqs. (\ref{LO_pseudo}) and derive          
the ``partially improved'' results on LO valence helicity PDFs
$\Delta {\hat q}_V(10\text{ GeV}^2,x)$ solving the system (\ref{LO_pseudo}) at $Q^2_{ref}=10\text{ GeV}^2$ with
the improved LO parametrization on $q_V$ (bold solid lines in Fig. \ref{fig:x_lo_uv_Q2_10}) in denominators instead of 
the original LO parametrization.

\begin{figure}
\center
\includegraphics[height=6.0cm]{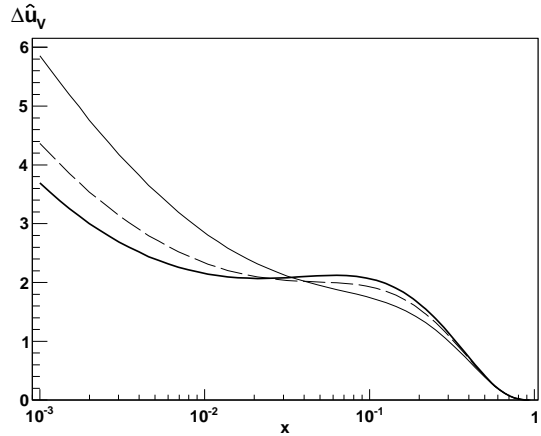}
\caption{
\footnotesize
$\Delta {\hat u}_V$ at  $Q_{ref}^2=10\text{ GeV}^2$ versus $x$.
Bold solid line corresponds to improved LO parametrizations on $\Delta {\hat q}_V$, which is obtained
from DSSV NLO parametrization via inverse connection formula (\ref{NLOtoLO_pol}).
Thin solid line corresponds to direct 
solution of the system (\ref{LO_pseudo}) with respect to $\Delta {\hat q}_V$,
where asymmetries $A_{p,(d)}^{``exp"}{\bigl |}_Z$ are taken at $Q_{ref}^2=10\text{ GeV}^2$ and MSTW2008 LO parametrization
on  ${\hat q}_V$ is used in denominators.
Asymmetries  $A_{p,(d)}^{``exp"}{\bigl |}_Z$ are calculated substituting  
DSSV NLO parametrization of polarized PDFs,  
MSTW2008 NLO parametrization of unpolarized PDFs, 
and AKK parametrization of FFs 
into the respective NLO equations (Eqs. (6-10) in Ref. \cite{shev}).
Long-dashed line (``partially improved'' LO result on  $\Delta {\hat q}_V$) is obtained in the same way, but
with the application of the improved MSTW2008 LO parametrization
of ${\hat q}_V$ (bold solid lines in Fig. \ref{fig:x_lo_uv_Q2_10}) 
in denominators of  (\ref{LO_pseudo}) instead of the original  MSTW2008 LO parametrization.
}
\label{fig:dssv_lo_qv_3}
\end{figure}

The 
results on ``partially improved'' $\Delta {\hat u}_V$ (long-dashed line) 
are presented in Fig. \ref{fig:dssv_lo_qv_3}
in comparison with the respective ``naive'' (thin solid line, the same as in Figs. \ref{fig:dssv_lo_qv}, \ref{fig:dssv_lo_qv_plus_our} 
) 
and ``completely improved'' (bold solid line, the same as in Fig. \ref{fig:dssv_lo_qv_plus_our})
LO results.
One can see that impact of the ``partial improvement'' on  $\Delta{\hat u}_V$ is very significant, and
the respective curve is even much closer to the line corresponding to the
``completely improved'' LO results. The picture for  $\Delta{\hat d}_V$ and for
application of GRSV NLO parametrization in l.h.s. of Eq.(\ref{LO_pseudo}) is absolutely analogous. 
 
Thus, one can hope that even in the cases when the state of art 
still 
does not allow to extract PDFs/FFs  
from the data at NLO (for instance, the TMDs case), it is 
possible, nevertheless, to significantly improve
the respective LO results only due to application of the improved LO unpolarized PDFs in denominators of the
asymmetries analyzed at LO in standard way.

{\it In summary}, 
the non singlet (inverse) connection formulas (\ref{NLOtoLO}), (\ref{NLOtoLO_pol}) between LO and NLO QCD orders were applied
to both unpolarized and polarized valence PDFs. 
It was shown that the connection formulas produce
improved LO results on valence PDFs 
without exclusion from the analysis the data coming from the small and moderate $Q^2$ region
(only the standard cut $Q^2>1\text{ GeV}^2$ should be applied).
The respective improved LO parametrizations on the valence PDFs were obtained (bold solid lines in Figs. \ref{fig:x_lo_uv_Q2_10}, \ref{fig:dssv_lo_qv_plus_our}),
and they strongly differ from the respective original LO
parametrizations, obtained by using the most of data 
produced at small and moderate $Q^2$ values.
Of importance is the argued statement that within the standard procedures of LO analysis (LO theoretical expressions
are applied to the measured asymmetries and cross sections) the same improved LO results 
could be obtained by using only the data produced  
at extremely high $Q^2$ values, that is hardly possible in reality.
It was argued that even in the cases when the state of art 
still 
does not allow to extract PDFs/FFs  
from the data at NLO (for instance, the TMDs case), it is 
possible, nevertheless, to significantly improve
the respective LO results only due to application of the improved LO unpolarized PDFs in denominators of the
asymmetries analyzed at LO in the standard way.

\end{document}